\documentstyle[11pt,aaspp4,psfig]{article}  

\def\etal{{\it et~al.~}}
\def\bsax{{\it Beppo-SAX~}}
\def\ginga{{\it Ginga~}}
\def\einstein{{\it Einstein~}}
\def\exosat{{\it EXOSAT~}}

\def\rosat{{\it ROSAT~}}

\def\ph{$~ph~cm^{-2}~s^{-1}~keV^{-1}~$}
\def\erg{$~erg~cm^{-2}~s^{-1}~$}
\begin{document}

\newcommand{\lessim}{\ \raise -2.truept\hbox{\rlap{\hbox{$\sim$}}\raise5.truept
	\hbox{$<$}\ }}			

\title{Hard X-ray Radiation in the Coma Cluster Spectrum}

\author{Roberto Fusco-Femiano} 
\affil{Istituto di Astrofisica Spaziale, C.N.R., via del Fosso del Cavaliere, 
I-00133 Roma, Italy - dario@saturn.ias.rm.cnr.it}
\author{Daniele Dal Fiume}
\affil{TESRE, C.N.R., via Gobetti 101, I-40129 Bologna, Italy}
\author{Luigina Feretti}
\affil{Istituto di Radioastronomia, C.N.R., via Gobetti 101, I-40129 Bologna, 
Italy}
\author{Gabriele Giovannini}
\affil{Istituto di Radioastronomia, C.N.R., via Gobetti 101, I-40129 Bologna, 
Italy}
\affil{Dip. di Fisica, Univ. Bologna, via B. Pichat 6/2, I-40127 Bologna, Italy}
\author{Paola Grandi}
\affil{Istituto di Astrofisica Spaziale, C.N.R., via del Fosso del Cavaliere,
I-00133 Roma, Italy}
\author{Giorgio Matt}
\affil{Dip. Fisica, Univ. ``Roma Tre", via della Vasca Navale 84, 
I-00146 Roma, Italy}
\author{Silvano Molendi}
\affil{Istituto di Fisica Cosmica, C.N.R., via Bassini 15, I-20133 Milano, 
Italy}
\author{Andrea Santangelo}
\affil{Istituto di Fisica Cosmica ed Applicazioni all'Informatica, C.N.R., 
via U. La Malfa 153, I-90146, Palermo, Italy}

\begin{abstract}
Hard X-ray radiation has been detected for the first time in the Coma cluster
by \bsax.
Thanks to the unprecedented sensitivity of the 
Phoswich Detection System (PDS) instrument, 
the source has been detected up to $\sim$80 keV. 
There is clear evidence (4.5$\sigma$) 
for non-thermal emission in excess of thermal 
above $\sim$25 keV. 
The hard excess is very unlikely 
due to X Comae, the Seyfert 1 galaxy present in the field of view of the PDS.
 
A hard spectral tail due to inverse Compton on CMB photons is 
predicted in clusters,
 like Coma,  with radio halos. Combining the present results with radio
observations, a volume-averaged intracluster magnetic
field of $\sim 0.15\mu G$ is derived, 
while the electron energy density of the emitting electrons
is $\sim 7\times 10^{-14}~erg/cm^3$.  
\end{abstract}

\keywords{cosmic microwave background --- galaxies: clusters: individual (Coma) --- magnetic fields --- radiation mechanisms: non-thermal --- X-rays: galaxies} 

\section{ Introduction}

Non-thermal hard X-ray radiation (HXR) is predicted
in galaxy clusters with radio haloes, such as Coma C in the central
region of the Coma cluster (Willson 1970), due
to inverse Compton (IC)
scattering by relativistic electrons of the Cosmic Microwave Background (CMB)
photons.  
The combined radio and HXR detections directly yield an estimate of
the mean volume-averaged  
intracluster magnetic field, B, and of the electron energy density, $\rho_e$.
Actually, to obtain informations on $\rho_e$, knowledge of
the size of the radio source and of the distance of the cluster are also needed.
It is worth remarking that the procedure is 
based essentially only on observables.
An accurate estimate of magnetic fields and electron and proton densities
present in clusters of galaxies is essential for a clearer picture of the
intracluster environment. In particular, these quantities are 
crucial to establish the role of
magnetic fields on the cluster dynamical history and the contribution of the
relativistic particles to the intracluster gas heating (Rephaeli 1979).

Here we present the first detection of hard X-ray emission from the 
Coma Cluster, obtained by \bsax (Boella \etal 1997).
Preliminary results have already been presented by
Fusco-Femiano \etal (1998). 
The Coma cluster was observed in December 1997
for an exposure time of $\sim 91$ ksec. Here we discuss results from the 
two passively
collimated instruments,
the High Pressure Gas Scintillation Proportional Counter (HPGSPC),
whose energy range is 4-120 keV, and the Phoswich Detection System (PDS),
working in the 15-200 keV energy range.
 
The main aim of this long observation was to search for
non-thermal hard X-ray radiation exploiting the unique capabilities
of the PDS :
an overall sensitivity better than a few times $10^{-6}$\ph
in the 40-80 keV energy band, a relatively small field of view
(FWHM=$1.3^{\circ}$, hexagonal), a
 wide 15-200 keV energy range, a low and stable background thanks to the
equatorial orbit.

The cluster was also observed with the Medium Energy
Concentrator Spectrometer (MECS), an imaging instrument
working in the 1.5-10 keV energy range. Here we briefly summarize the results 
that will be discussed in detail in a forthcoming paper.
Spatially resolved spectroscopy was performed on the core of the
cluster with concentric annuli of $2^{\prime}$ within a region
$8^{\prime}$ in radius ($\sim$325 kpc). There is no evidence 
for variations of the intracluster medium (ICM) temperature and
metallicity in the observed region. The average temperature was determined 
to be
$9.1\pm0.1$ keV, while the average iron 
abundance is 0.26$\pm$0.02.   

Throughout the paper
we assume a Hubble constant of $H_o = 50~km~s^{-1}~Mpc^{-1}~h_{50}$ and
$q_0 = 1/2$, so that an angular 
distance of $1^{\prime}$ corresponds to 40.6 kpc ($z_{Coma} = 0.0232$). 
Quoted confidence intervals are $68\%$ level, if not otherwise specified.

\section {PDS and HPGSPC Data Reduction}

The PDS and the HPGSPC instruments use the rocking
collimator technique for background subtraction with angles of $3.5^{\circ}$
 and
$3^{\circ}$, respectively. In the PDS, the standard observation strategy is to
observe the X-ray target with one collimator and to monitor
the background level on both sides of the source position with the other,
so that a continuous monitoring of the source and background is obtained. The
standard dwell 
time in each position of both collimators is 96 sec. 
The background level of the PDS is the lowest obtained thus far with high
energy instruments onboard satellites, thanks to the equatorial orbit.
In the 15-300 keV energy band, the background is 
$\sim 2\times 10^{-4}~cts~cm^{-2}~s^{-1}~
keV^{-1}~$  
(Frontera \etal 1997). The background is very stable, again thanks to the
favourable orbit, and no modelling of the time variation of the
background is required.  
Counts are excluded from the analysis for 300 sec after the passage of 
the satellite
through the trapped radiation of the South Atlantic Geomagnetic Anomaly. 
In the HPGSPC the collimator assumes on and off-source position with the
same dwell time of the PDS (Manzo \etal 1997).

The correctness of the PDS background subtraction has been checked by 
verifying 
that the counts fluctuate about zero flux as the signal falls below
detectability. This happens
 at energies greater than $\sim$80 keV. Two broad strong features 
peaked at $\sim$30 keV and $\sim$60 keV
are present in the PDS background spectrum (Frontera \etal 1997). However,
even excluding the counts in the energy bands of these features, 
hard X-ray radiation is still detectable.
\begin{figure}[ht]
\centerline{\psfig{figure=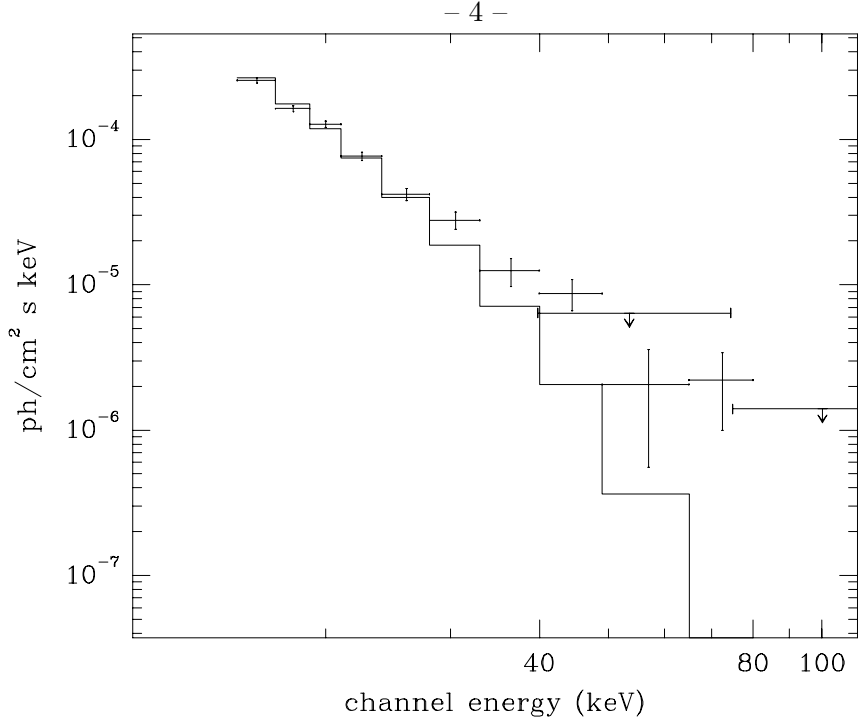,height=9.5cm,width=9.5cm}}
\vskip20pt
\caption{PDS data. The continuous line represents the best fit
with a thermal component at the average cluster gas temperature of 8.21 keV
(Hughes \etal 1993). The upper limits refer to the OSSE experiment (Rephaeli,
Ulmer \& Gruber 1994).}
\end{figure}
\section {PDS and HPGSPC Data Analysis and Results}

The PDS detected hard X-ray emission in the 15-80 keV energy range in the
Coma cluster spectrum, as shown in Fig. 1.
We started the spectral
analysis by working only with the PDS data to have a robust indication of the
presence of a non-thermal component in the observed HXR radiation. 
The MECS data are not very helpful in this respect, as
its field of view is much smaller than
that of the PDS. 
With the PDS data alone, however, it is not possible to perform complex
spectral fits, and one has to resort to fixing the temperature of the
intracluster medium.
Different temperatures, in the range 7.5-8.5 keV, have been
reported by previous X-ray observatories for the Coma cluster 
(Hughes \etal 1988; Hughes, Gorenstein 
\& Fabricant 1988; Watt \etal 1992; Hughes \etal 1993), but with a 
clear pattern: the larger
the field of view, the lower the measured temperature.
By fixing the temperature of the thermal component to the average
cluster value of 8.21 keV,  as determined by 
the Large Area Counter (LAC) experiment onboard \ginga (whose field of 
view of $1^{\circ}\times 2^{\circ}$ is not much different from 
that of the PDS), we obtain the fit shown in figure 1. 
Adopting a temperature as high as
9.1 keV, as observed by the MECS in the cluster core, the results are
substantially the same. The flux of the thermal
component is $\sim 3.3\times 10^{-10}$\erg in
the 2-10 keV energy range; the 
reduced $\chi^2$ of the fit is 3.2 for 9 degrees of freedom, mainly 
due to a clear 
excess above
25 keV at a 4.5 $\sigma$ level. Leaving the temperature as free parameter,
a value of kT=10.73$^{+0.81}_{-0.74}$ is obtained; 
the reduced $\chi^2$ is 2.14 for 8 degrees of freedom. The $\chi^2$ value 
decreases significantly when a second component is added. Modelling this
further component as a power law, as for non-thermal emission,
the reduced $\chi^2$ value is 0.92 for 7 
degrees of freedom. The  
confidence interval for the power law 
spectral index is rather large, 0.7-3.6, but  
the flux ($\sim 2.2\times 10^{-11}$\erg
in the 20-80 keV energy range), does not depend much on 
the power law index. 
On the other hand, if we consider  
a second thermal component, instead of the non-thermal one,
the fit requires a temperature greater than 40 keV. This unrealistic value 
may be interpreted as a strong indication that the
detected hard excess is due to a non-thermal mechanism. It is worth remarking 
that the X-ray fluxes measured by
\ginga and PDS coincide, within the errors, in their common energy range
15-20 keV, where the flux  is still dominated by the ICM thermal component.
The contribution of the non-thermal component is not
greater
than 20$\%$ at 20 keV, even adopting the rather steep value of 2.34 
(see $\S$ 4)
for the spectral
index of the hard flux.
%
\begin{figure}[ht]
\centerline{\psfig{figure=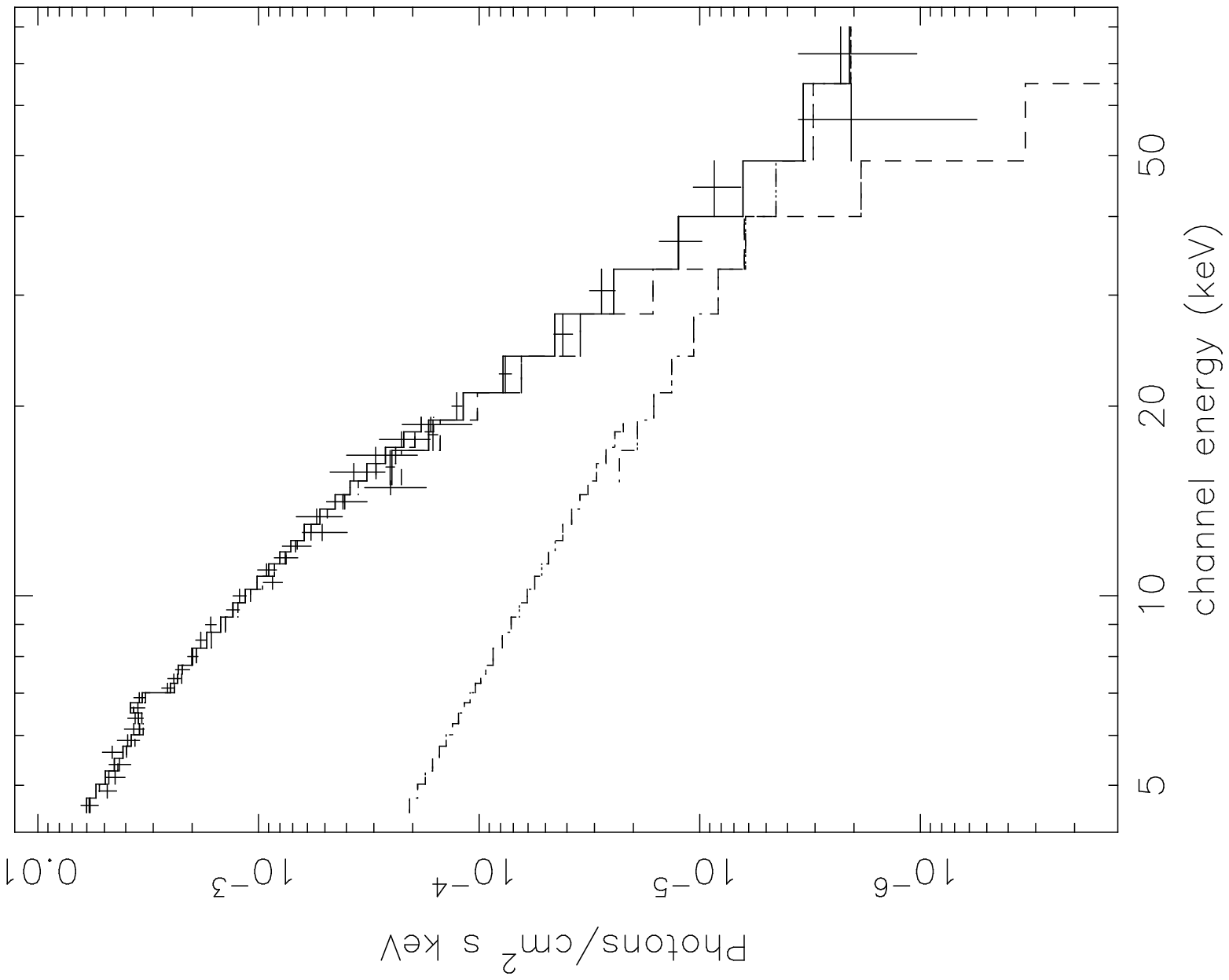,height=22.5cm,width=12.0cm,angle=-90}}
\caption{The continuous line is the best-fit to the
HPGSPC (4.5-20 keV) and PDS data (15-80 keV). The dashed line represents
the thermal component (kT=8.5$^{+0.6}_{-0.5}$ keV), while the dashed
straight line is
the non-thermal component with a spectral index of 1.6. A constant value of
0.90 takes into account the relative calibration of the two data sets. The
$\chi^2_{red}$ is 0.80 for 34 degrees of freedom.}
\end{figure}
The analysis of the PDS data alone cannot provide 
a precise estimate of the non-thermal spectral index. 
A tighter constraint can be obtained by including 
in our analysis the HPGSPC data in the 4.5-20 keV energy range, as shown 
in figure 2.
The HPGSPC instrument has a field of view ($1^{\circ}\times
1^{\circ}$), comparable to that of the PDS ($1.3^{\circ}$, hexagonal),
and can therefore be used for a global spectral fit. 
Combining HPGSPC and PDS data it is then possible to leave the temperature
of the intracluster medium as a free parameter in the fit procedure. 
The relative normalization of the two instruments has been left free, 
to account for the slightly different field of views and the
remaining uncertainties in the intercalibration.
We obtain an ICM temperature of 
8.5$^{+0.6}_{-0.5}~$keV, consistent with the \ginga determination, and an
iron abundance of $0.15^{+0.6}_{-0.4}$.
 The reduced $\chi^2$ is 0.80
for 34 degrees of freedom. The resulting allowed spectral index 
range is 0.7-2.5, smaller
than the previous one 
determined using only the PDS data, but still not sufficient
to distinguish between different emission mechanisms.
If we fix the parameters of the thermal component to their
best-fit values we obtain $1.57^{+0.36}_{-0.39}$ ($90\%$) for the photon
index of the non-thermal radiation, corresponding 
to a  contribution of
the non-thermal emission to the 2-10 keV flux between $\sim (2-10)\%$. 
   
\section{Discussion}
The PDS onboard the \bsax satellite detected hard X-ray emission up to
energies of $\sim$80 keV in the spectrum of the Coma cluster, with a clear 
excess above the thermal intracluster emission, which is best 
explained as a non-thermal component. 

One possible explanation for the observed excess is emission by a
different source in the field of view.
The most qualified candidate is the Seyfert 1 galaxy X Comae
at z=0.092$\pm$0.002 (Bond \& Sargent 1973). The \rosat PSPC observation
(Dow \& White 1995) reports a power law spectrum with photon index 
2.50$\pm$0.16 and a flux of 3.6$\times 10^{-12}$\erg in the 0.4-2.4 keV band
(corresponding to 
$\sim 1.6\times 10^{-12}$\erg in the range 2-10 keV). X Comae has been 
observed also by \exosat (Branduardi-Raymont \etal 1985) and by \einstein 
IPC Slew Survey (Elvis \etal 1992) at approximately the same flux level.  
The steep photon index reported by \rosat
requires an unusual variability of a factor $\sim$65 to explain
the detected hard excess, 
considering that the source is $\sim 27^{\prime}$
off-axis and $N_{\rm H, gal} = 9\times 10^{19}~cm^{-2}$. With a more typical
photon index of 1.8 the variability factor is $\sim$11, 
no longer extreme but still large. Luckily enough, X Comae is located just 
on the edge of the field of view of the MECS,
and part of the Point
Spread Function (PSF) of the source lies within the field of view 
of the detector.
Considering the location of X Comae and the lack of detection, it is
possible to estimate an upper limit to the flux of the source of the order
of $\sim 4\times 10^{-12}$\erg (2-10 keV), 
much lower than the 
flux of $\sim 2.9\times 10^{-11}$\erg (photon index = 1.8) required to 
account for the HXR.
 
If, instead,  we assume that the hard spectral tail detected at 
energies above 25 keV
is due to
relativistic electrons scattering the CMB photons, 
it is possible to derive, using only observables, the value of the 
mean volume-averaged intracluster magnetic
field, B. In the Coma cluster, the radio halo
Coma C has an extension of $\sim$1 Mpc in radius and is located in the central 
region of the cluster. 
Relating the observed synchrotron radio flux
to the X-ray IC flux we obtain:
$$f_x = C(\alpha)~F_r
B^{-(1+\alpha)}\nu_r^{\alpha}\epsilon_x^{-(1+\alpha)}$$

\noindent
where $\alpha$ is the slope of the radio halo spectrum and $F_r$ is the radio 
flux at the frequency $\nu_r$; for $C(\alpha)$ see
Rephaeli (1979).
\begin{figure}[ht]
\centerline{\psfig{figure=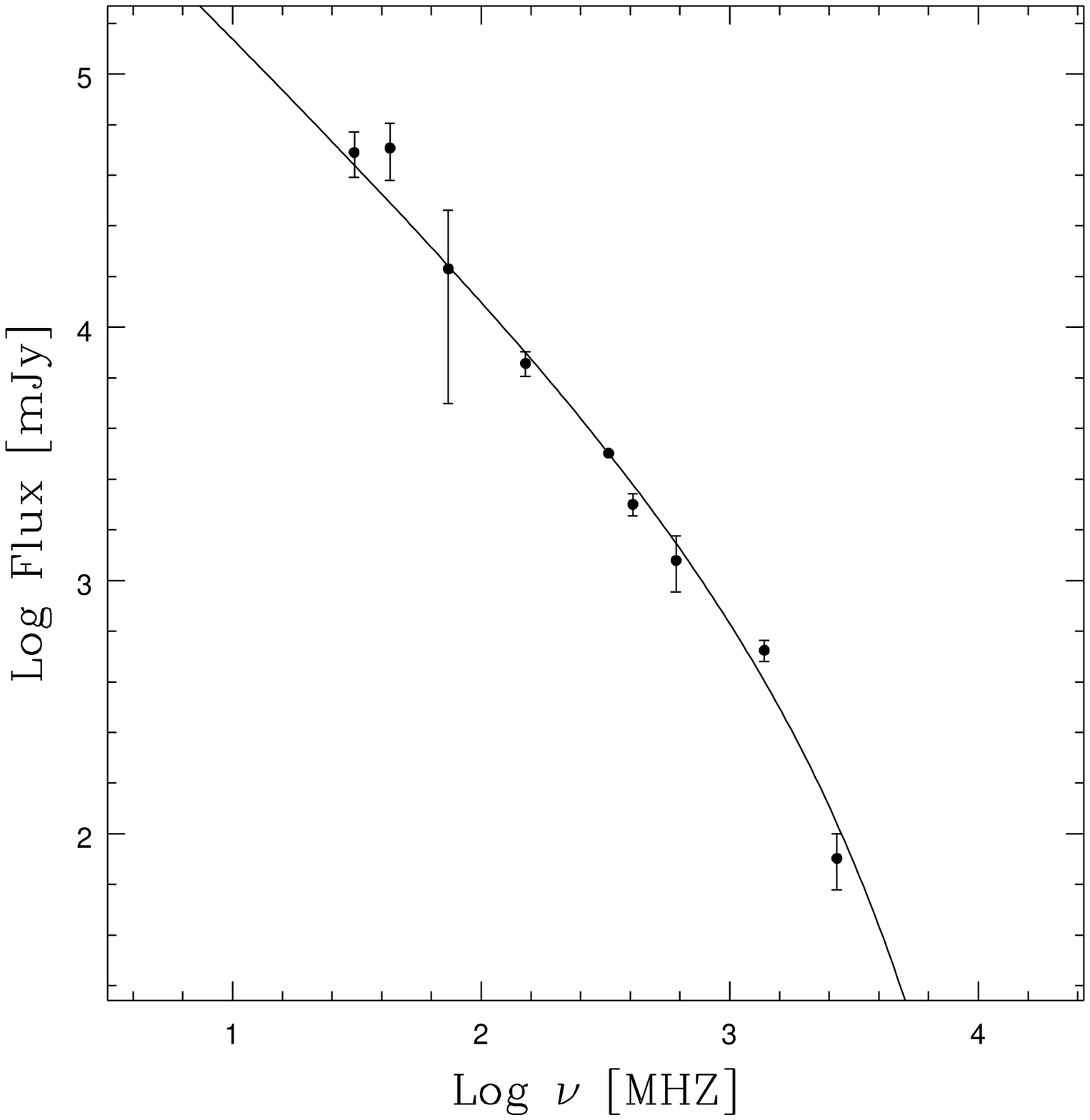,height=9.5cm,width=9.5cm}}
\caption{The radio halo spectrum of Coma C. The best-fit
is obtained taking into account the electron energy losses. The
$\chi^2_{red}$ is 2.78 for 6 degrees of freedom.}
\end{figure}
The value of the radio halo spectral index is still uncertain for Coma C.
Considering all the radio fluxes measured between 30.9 MHz and 1.4 GHz, 
Kim \etal (1990) and Giovannini \etal (1993) determined a value of 
$\alpha=1.34\pm 0.06$, while
$\alpha=1.16\pm 0.03$ is obtained using the flux measured at 1.4 GHz 
by Deiss \etal (1997). We have re-modeled the 
radio flux spectrum of Coma C between 30.9 MHz and 2.7 GHz using the
data set given in Giovannini \etal (1993), taking into account 
the electron energy losses due to
synchrotron and IC emission (see Fig. 3). The best-fit indicates an 
electron injection
spectrum of 2.92$\pm$0.20 ($\alpha=0.96\pm 0.10$, $90\%$), while 
curvature of the radio
spectrum is detected at $\nu_r >$ 100 MHz due to electron ageing. 
Considering that the electron energy range should be 2.4-4.9 GeV in order 
to give
 IC emission in the 20-80 keV band, and that the radio
spectrum begins to steepen at $\nu_r >$ 100 MHz, 
$\alpha$ is 0.96 for values of B$\leq 0.25\mu G$. The 
radio data and the 
X-ray spectrum detected by the PDS then imply a value of B$\sim 0.14~\mu G$,
while B$\sim 0.16~\mu G$ is obtained if the extreme value of
1.34 is adopted for $\alpha$. 
Using the size, R, and the distance, d, of the radio source it is possible
to estimate the electron energy density.
Assuming a radio
halo size R=1 Mpc ($\sim 25^{\prime}$) and a distance d=138 Mpc, the 
energy density of
electrons with energies $\geq$ 500 MeV is $\sim 7\times 10^{-14}~ ergs/cm^3$. 
The estimated value of B is not very different from the lower
limit of 0.1 $\mu G$ derived from the 2$\sigma$ upper bounds on the HXR
reported by the OSSE experiment (Rephaeli, Ulmer \& Gruber 1994). Binning the 
PDS data between $\sim$40 keV and
$\sim$80 keV,  
the HXR flux ($\sim 4\sigma$) is lower by a factor $\sim$2
with respect to the upper limit (see Fig. 1).

The non-thermal contribution to the 2-10 keV flux is 
$\sim 10\%,~\sim 16\%$ and
$\sim 24\%$ for $\alpha$= 0.96, 1.16 and 1.34, respectively. This contribution
has implications for the determination of the intracluster gas properties.

The value of the magnetic field derived here seems to be inconsistent
with the measurements of Faraday rotation of polarized radiation
through the hot ICM that give a line-of-sight value of B in the range
$\sim 2-6~h_{50}^{1/2}~\mu G$ (Kim \etal 1990; Feretti \etal 1995).
We note, however, that Feretti \etal (1995) inferred also the existence
of a weaker magnetic field component, ordered on a scale of about
a cluster core radius, with a line-of-sight strength in the range 
$\sim 0.1-0.2~h_{50}^{1/2}~\mu G$.
 From the results obtained here with the IC model, we can argue that the
strongly tangled magnetic field component of 6 $\mu G$ is likely present
in local cluster regions, while the overall cluster
magnetic field may be more reasonably represented by the weaker and
ordered component, whose strength is in good agreement with the
present estimate. Other determinations (or lower limits) of B based on
different methods are in the range 0.2-0.4$\mu G$  
(Hwang 1997;  
Bowyer \& Bergh\"{o}fer 1998; 
Sreekumar \etal 1996; 
Henriksen 1998). In particular, the equipartition magnetic field is 
$\sim 0.4~h_{50}^{2/7}~\mu G$ 
(Giovannini \etal 1993). 
Very attractive suggestions have been proposed regarding the origin of the
IC magnetic field, such as that in which galaxy motion may drive a turbulent 
dynamo which
amplifies faint seed fields to an average value of $\sim 0.1-0.2\mu G$.
 The seed fields are supplied through gas loss by galaxies (Goldman \& Rephaeli
1991; De Young 1992). Larger magnetic fields require different amplification
mechanisms, such as merger activity .

Different interpretations  have also been 
proposed to explain the non-thermal hard X--ray emission. 
Ensslin, Lieu \&
Biermann (1998) suggest that it might be 
bremsstrahlung emission by suprathermal electrons in the ICM, accelerated
by turbulences within the medium. 
Another proposed emission mechanism is given by IC 
scattering of a 
large population of cosmic rays by
the CMB photons (Lieu \etal 1998b). In this model the relativistic component 
could be also responsible
for the soft excess discovered in a few clusters in the 
69 eV - 0.4 keV energy band (Lieu \etal 1996a,b; Bower, Lampton \& Lieu 1996; 
Fabian 1996; 
Mittaz, Lieu \& Lockman 1998; Bowyer, Lieu \& Mittaz 1998; Kaastra 1998). In
particular, the re-modeled EUV and
soft X-ray data by Lieu \etal (1998a) lead to a photon index of 
$\sim$1.75. An extrapolation
of the EUV spectra to higher energies gives an excess flux between 20-80 keV
of $\sim 1\times 10^{-11}$\erg, only a factor $\sim$2 lower than 
the PDS excess, suggesting a possible physical connection between the
soft and hard excesses.

The next step in studying the hard X--ray emission would be a precise
estimate of its spectral shape, which would help in discriminating 
between competing emission mechanisms. A longer \bsax observation of the
Coma cluster would be valuable in this respect. 
Sarazin \& Lieu (1988) suggest
that a relic population of very low energy cosmic ray electrons may be
responsible for the
IC EUV excess detected also in clusters of galaxies in which radio haloes
are absent. 
It would be of great interest to investigate radio quiet clusters of
galaxies to verify whether hard X-ray excesses are 
common in galaxy clusters. 
 
\acknowledgments 

We thank F. Fiore, P. Giommi and L.Piro for useful suggestions 
regarding data
analysis, the referee for valuable comments and suggestions and W.D. Cotton
for a critical reading of the manuscript.

\end{document}